\documentclass[conference]{IEEEtran}
\IEEEoverridecommandlockouts
\usepackage{cite}
\usepackage{amsmath,amssymb,amsfonts}
\usepackage{algorithmicx}
\usepackage{graphicx}
\usepackage{textcomp}
\usepackage{xcolor}
\usepackage{cite}
\usepackage{mdwmath}
\usepackage{mdwtab}
\usepackage{mathtools}
\usepackage{eqparbox}
\usepackage[tight,footnotesize]{subfigure}
\usepackage[rightcaption]{sidecap}
\usepackage{graphicx} 
\graphicspath{{images/}}
\usepackage{url}
\usepackage{lettrine}
\usepackage{color}
\usepackage{verbatim}
\usepackage{amsmath,amssymb,amsfonts}
\usepackage{graphicx}
\usepackage{textcomp}
\usepackage{xcolor}
\usepackage{subfig}
\usepackage[ruled,vlined]{algorithm2e}
\usepackage{blindtext}
\newtheorem{definition}{Definition}[section]
\def\BibTeX{{\rm B\kern-.05em{\sc i\kern-.025em b}\kern-.08em
    T\kern-.1667em\lower.7ex\hbox{E}\kern-.125emX}}
\begin{document}

\title{A Distributed Sparse Channel Estimation Technique for mmWave Massive MIMO Systems}

\author{
	\IEEEauthorblockN{Maria Trigka, Christos Mavrokefalidis, Kostas Berberidis}
	\\
	\IEEEauthorblockA{Department of Computer Engineering and Informatics, University of Patras, Greece \\ \{trigka,maurokef,berberid\}@ceid.upatras.gr}
}

\maketitle

\begin{abstract}
In this paper, we study the problem of sparse channel estimation via a collaborative and 
fully distributed approach. The estimation problem is formulated in the angular domain by exploiting the spatially common sparsity structure of the involved channels in a multi-user scenario. The sparse channel estimation problem is solved via an efficient distributed approach in which the participating users collaboratively estimate their channel sparsity support sets, before locally estimate the channel values, under the assumption that global and common support subsets are present. The performance of the proposed algorithm, named WDiOMP, is compared to DiOMP, local OMP and a centralized solution based on SOMP, in terms of the support set recovery error under various experimental scenarios. The efficacy of WDiOMP is demonstrated even in the case in which the underlining sparsity structure is unknown.
\end{abstract}


\section{Introduction}

In 5G (and beyond) wireless communication systems, the millimeter wave (mmWave) spectrum is considered for increasing, among others, the transmission capacity \cite{alkhateeb2014mimo},\cite{zhang2015massive},\cite{heath2016overview}. Signals in the mmWave bands, however, experience high sensitivity to path loss, limiting the communication range. To mitigate this loss, massive Multi-Input, Multi-Output (MIMO) and beamforming have been suggested in order to guarantee reliable communication. To benefit from the massive MIMO array gain and perform high directional beamforming, channel information is required between all antenna pairs, increasing the training overhead. On the other hand, due to limited scattering  \cite{rao2014distributed}, a small number of significant paths actually contribute to the transmission. Thus, in the angular domain, channel estimation can be achieved with reduced training overhead by identifying the channel gains, the Angles of Arrival (AoA) and Angles of Departure (AoD), related only to those paths via Compressed Sensing (CS) \cite{alkhateeb2014channel}. 

Exploiting CS for channel estimation has been an active research area in recent years \cite{bajwa2010compressed}. Most of the proposed works employ CS algorithms either at individual nodes \cite{alkhateeb2014channel} or at a central point (namely, a Base Station - BS), in which multiple measurements of groups of nodes are fed back \cite{rao2014distributed}, \cite{gao2015spatially}. In the latter case, the CS algorithms are able to exploit common sparsity patterns that manifest in the node measurements due to the transmission environment and, thus, achieve improved estimation performance or reduced training overhead. In this paper, a \textit{fully} distributed channel estimation algorithm is proposed that (a) exploits common sparsity patterns among the collaborating nodes for improving performance and (b) requires no central point for gathering measurements and performing the associated processing. The deployment of the proposed algorithm is able to reduce the channel uses for training by the BS and improve the channel estimation performance of the nodes, compared to the case they operate as individual entities (especially, at the low SNR regime). 

In more detail, in this paper, it is assumed that the collaborating nodes receive, in the downlink, a number of paths from identical angles while, in the remaining ones, the angles might be different \cite{rao2014distributed}. This means that the involved channels have a global and, possibly, common sparsity patterns (or support sets), observed by all and groups of nodes, respectively. Capitalizing on the distributed sparse coding literature, where the previous ``sparsity model'' was introduced \cite{baron2006distributed} and solved \cite{wimalajeewa2014omp}, \cite{sundman2014distributed}, the proposed technique extends the Distributed Orthogonal Matching Pursuit (DiOMP) algorithm \cite{sundman2014distributed} by employing a weighted majority voting mechanism with adaptive combination weights to identify the desired support sets. The voting mechanism, contrary to the simple majority voting rule used in \cite{sundman2014distributed}, assumes no previous knowledge about the structure of the global and common sparsity patterns (which in the distributed literature answers also to the term ``node-specific'' \cite{Chaves2015}), apart from the total number of involved paths. The performance of the proposed algorithm is extensively assessed via simulations in three different scenarios and it turns out that the proposed scheme exhibits considerable improvement over the existing ones. 

The rest of this paper is organized as follows. In Section \ref{sec:signal_model}, the data model is described. In Section \ref{sec:smodel2}, the proposed distributed MIMO downlink channel modeling and estimation, with the aid of the CS theory, is elaborated in detail. Finally, in Section \ref{sec:res}, several experiments are presented to verify the efficiency of the proposed approach.

\section{System Model}\label{sec:signal_model}
In this section, a brief description of the adopted data model is provided. Consider a massive MIMO system with $K$ single antenna users and a BS equipped with $N$ antennas which broadcasts a sequence of $T$ training pilot vectors, denoted by $\boldsymbol{X} \in \mathbb{C}^{T \times N}$, towards each user for estimating the downlink channel. 
Then, the downlink received signal $\boldsymbol{y}_k \in \mathbb{C}^{T \times 1}$ at the $k$-th user is given by
\begin{equation}
\boldsymbol{y}_k = \boldsymbol{X}\boldsymbol{h}_k + \boldsymbol{n}_k,
\label{eq:yk}
\end{equation}
where $\boldsymbol{n}_k \in \mathbb{C}^{\textrm{T}\times1}$ stands for additive noise which is modeled as random vector with elements being independent and identically distributed as complex Gaussian random variables with zero mean and variance equal to $\sigma^2$. 

Assuming a block, flat fading channel $\boldsymbol{h}_k\in \mathbb{C}^{N\times 1}$ from the BS to the $k$-th user, the Geometry-Based Stochastic Channel Model (GSCM) \cite{ding2018dictionary} is adopted which, after dropping the index $k$ for clarity, can be written as
\begin{equation}
\boldsymbol{h} = \sum_{c=1}^{N_c}\sum_{s=1}^{N_{c,s}} \gamma_{c,s} \boldsymbol{a}(\theta_{c,s}),
\label{eq:hk1}
\end{equation}
where $N_c$ and $N_{c,s}$ stand for the number of scattering clusters and the number of sub-paths per scattering cluster, respectively, $\gamma_{c,s}$ is the complex gain of the $s$-th sub-path in the $c$-th scattering cluster, $\theta_{c,s}$ is the corresponding AoD and 
\begin{equation}
\boldsymbol{a}(\theta_{c,s})=\frac{1}{\sqrt{N}}\left[1,\, e^{j\frac{2d\pi}{\lambda_c}\sin{\theta_{c,s}}},\dots, e^{j\frac{2(N-1)d\pi}{\lambda_c}\sin{\theta_{c,s}}}\right]^\textrm{T}
\end{equation}
is a steering vector, assuming a Uniform Linear Array (ULA).
From now on, we denote the true AoDs as $\theta_l, l = 1,2,\hdots,L$, with $L = N_cN_{c,s}$. 

Exploiting the limited number of scattering clusters in the BS side and few active transmit directions for each user, the channel can be approximated by $\boldsymbol{h} \approx \boldsymbol{\Psi}\boldsymbol{w}$, where $\boldsymbol{w}$ is a sparse vector and $\boldsymbol{\Psi} \in \mathbb{C}^{N\times \hat{L}}$ is a sparsifying dictionary \cite{ding2018dictionary}. A typical $\boldsymbol{\Psi}$ is the normalized square DFT matrix, leading to the known ``virtual channel model'', which maps the spatial channel response to the angular domain. Alternatively, an overcomplete dictionary $\boldsymbol{\Psi}$ can be employed via the overcomplete DFT matrix, where $\hat{L}\gg N$ is the number of atoms in $\boldsymbol{\Psi}$ corresponding to the AoDs in a finer grid. In this case, $\boldsymbol{w} \in \mathbb{C}^{\hat{L}\times1}$ is able to approximate more closely the true angles at $\theta^l$. Thus, $\boldsymbol{y}_k$ can be written as
\begin{eqnarray}
\boldsymbol{y}_k \approx \boldsymbol{X}\boldsymbol{\Psi}\boldsymbol{w}_k + \boldsymbol{n}_k,
\label{eq:ykn}
\end{eqnarray}
where $\boldsymbol{X}\boldsymbol{\Psi}$ is the measurement matrix.
To achieve the CS recovery, the measurement matrix $\boldsymbol{X}\boldsymbol{\Psi}$ should satisfy specific conditions like the ones determined by the Restricted Isometry Property (RIP). For example, this can be achieved if the training matrix $\boldsymbol{X}$ has random elements which are identically and independently distributed
 \cite{ding2018dictionary}.
\section{Cooperative and distributed sparse channel estimation}\label{sec:smodel2}

In this section, the problem of mmWave MIMO sparse channel estimation is discussed. First, the adopted, joint sparsity model for the involved channels will be presented. Then, the proposed cooperative and distributed algorithm will be described. 

\subsection{Sparse Channel Modeling}
\begin{figure}
 \centering
 \includegraphics[scale = 0.5]{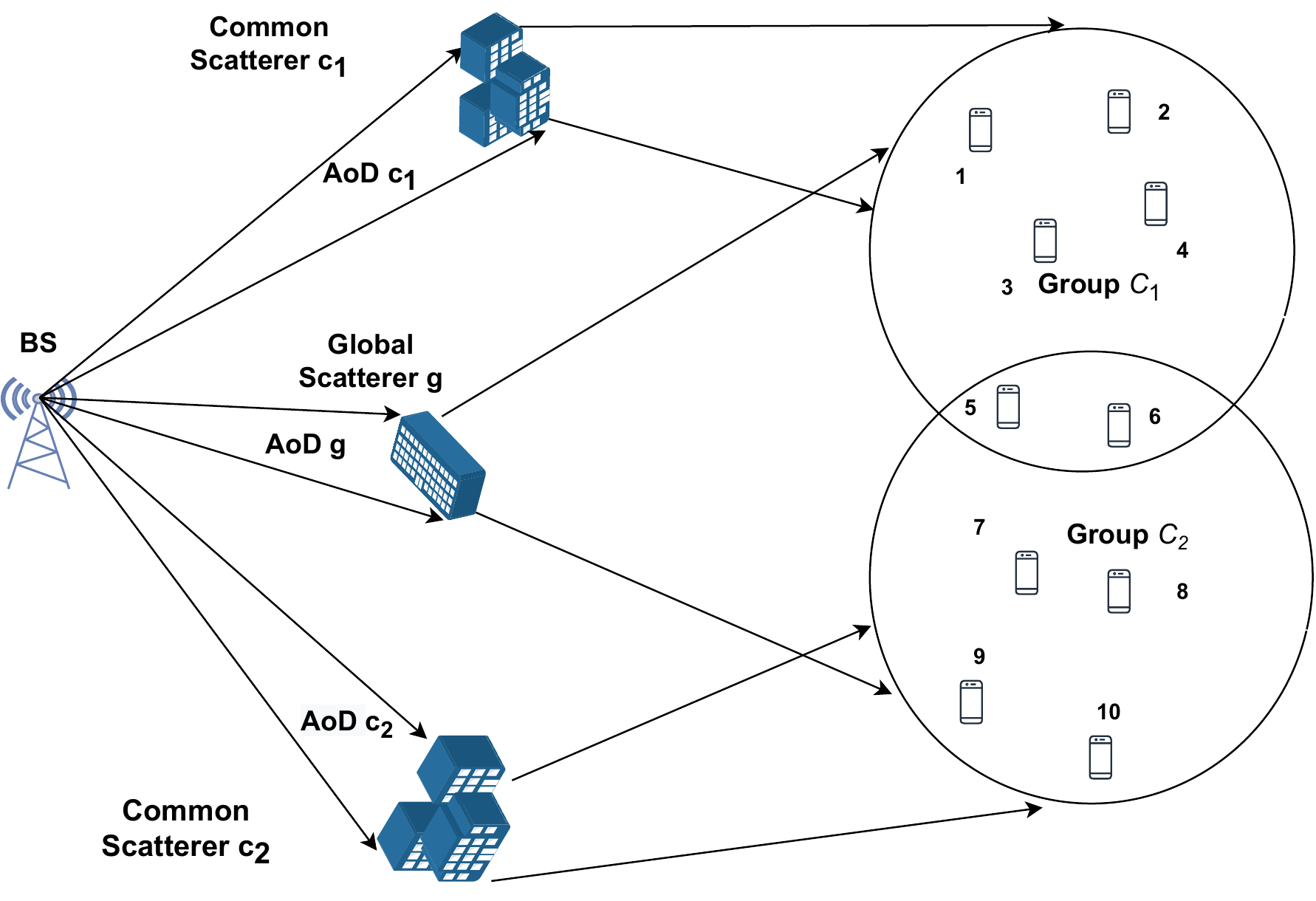}\hfill
 \caption{Downlink transmission scenario: $\mathcal{C}_1 = \{1,2,3,4,5,6\}$, $\mathcal{C}_2 = \{5,6,7,8,9,10\}$.} 
 \label{fig:net}
 \end{figure}

Let us consider a set $\mathcal{V} =\{1,2,\hdots,K\}$ of $K$ geographically distributed users with each one cooperating with all its connected neighbors. Due to the scattering nature of the environment (see Fig.~\ref{fig:net}), the users belong to groups depending on which scatterers are involved to their channels. Following \cite{rao2014distributed}, the users of a group face the same scattering structure because they are affected by the same scatterers, which means that the involved sparse vectors $\boldsymbol{w}$'s have the same sparsity support sets. However, the values of the coefficients at the corresponding positions may be different, due to the surrounding environment of each user. Between two groups, the scattering structure might be similar, namely, part of the sparsity support sets of the involved $\boldsymbol{w}$'s can be the same.

The performance of the downlink sparse channel estimation can be improved if we exploit the aforementioned relation on the sparsity support sets among the involved channels of the users \cite{rao2014distributed}, \cite{ gao2015spatially}. 
In more detail, the transmission environment at the BS side, as illustrated in Fig.\ref{fig:net}, can be organized into global and common (denoted as $g$ and $c_j$, respectively) scattering clusters, leading to respective AoDs that are global to all users, while others are only common to particular groups. For each user $k$, there is a subset $\boldsymbol{I}_k$ of indices $j = \{1,2,\hdots,J\}$ indicating the corresponding common interest support sets and groups in the network.
This situation translates to the respective atoms of the employed dictionary $\boldsymbol{\Psi}$ which can be distinguished as global (to all users) and common (to groups of users), as well. 

Adopting the aforementioned sparsity model, the sparse representation vector $\boldsymbol{w}_k$ of user $k$ can be formulated as 
\begin{equation}
    \boldsymbol{w}_k = \boldsymbol{w}_k^g + \sum_{j \in I_k} \boldsymbol{w}_k^{c_j},
\end{equation}
where $\boldsymbol{w}_k^g$ stands for the sparse representation vector whose support set is globally shared among all users in the network and $\boldsymbol{w}_k^{c_j}$ stands for the sparse representation vector whose support set is commonly shared among users in a particular group in the network. Hence, a \textit{Global and Multiple Common} support sets model arises, which captures the sparse properties of the involved channels among the BS and each user in the network.

Moreover, it is assumed that $\boldsymbol{w}_k^{g}$ of each user $k$ consists of $L_g$ non-zero parameters and $\boldsymbol{w}_k^{c_j}$'s consist of $L_{c_j}$ non-zero parameters. Hence, the sparsity profile of a user $k$ consists of global and multiple common interest support subsets. In the following, some definitions will be provided. In the following, some definitions will be provided. There, $\Omega = \{1,2,\hdots,\hat{L}\}$ denotes the set of atom indices. 
\begin{definition}
	(Global Interest Support Set): Let the globally shared sparse representation vector $\boldsymbol{w}^{g}_k$ with only $L_{g}$ nonzero entries. The global interest support set is defined as 
	\begin{equation}
	\mathcal{S}^{g} = supp(\boldsymbol{w}^{g}_m) = supp(\boldsymbol{w}^{g}_n),\forall m,n \in \mathcal{V},
	\end{equation}
where $\mathcal{S}^{g} \subset \Omega$ and $|\mathcal{S}^{g}| = L_{g}$ ($l_0$ sparsity). 
\end{definition}
\begin{definition}
	(Common Interest Support Set): Let the sparse representation vector $\boldsymbol{w}^{c_j}_k$ with only $L_{c_j}$ nonzero entries. The common interest support set is defined as 
	\begin{equation}
	\mathcal{S}^{c_j} = supp(\boldsymbol{w}^{c_j}_m) = supp(\boldsymbol{w}^{c_j}_n), \forall m,n \in \mathcal{C}_j,
	\end{equation}
where $\mathcal{S}^{c_j} \subset \Omega$,  $|\mathcal{S}^{c_j}| = L_{c_j}$ ($l_0$ sparsity) and $\mathcal{C}_j$ a set of users indices that are impacted by the common scattering cluster $c_j$, for $j=1,2,\hdots,J$, and are concerned about $\mathcal{S}^{c_j}$.
\end{definition} 

Therefore, the support set of a user $k$ can be formulated as $\mathcal{S}_k = \mathcal{S}_k^{g} \cup \{\mathcal{S}_k^{c_j}\}$, for all $j \in I_k$, meaning that the joint sparsity $L \le L_{g} + \sum_{j \in I_k} L_{c_j}$. Here, although $\mathcal{S}^{g}$ is the same for all users in the network or $\mathcal{S}^{c_j}$ is similar among users in the same sub-group $\mathcal{C}_j$, the corresponding non-zero values of $\boldsymbol{w}_k^{g}$ and $\boldsymbol{w}_k^{c_j}$ are still individual and possibly independent among the users. 

In summary, the channel $\boldsymbol{h}_k$ of user $k$ is written as 
\begin{equation}
\boldsymbol{h}_k = \boldsymbol{\Psi}(\boldsymbol{w}_k^{g} + \sum_{j \in I_k}\boldsymbol{w}_k^{c_j}).
\label{eq:channel}
\end{equation}

Considering (\ref{eq:channel}), our aim is to exploit the joint sparsity of the support sets of neighboring users to estimate the per user sparse channel in a cooperative and distributed way. Such an approach will be elaborated in the following.

\subsection{Proposed cooperative and distributed algorithm} \label{sec:CODIC}
The problem of estimating $\boldsymbol{h}_k$ in a collaborative and distributed manner, using the predefined dictionary $\boldsymbol{\Psi}$, is cast to the estimation of $\boldsymbol{w}_k$. The algorithm estimates collaboratively the support sets and locally, via Least-Squares (LS), the non-zero values at each user.

In more detail, the algorithm is based on DiOMP \cite{sundman2014distributed} which is improved in a key aspect by considering a new, weighted majority voting mechanism with adaptive weights, instead of the simple majority voting that DiOMP considers. In particular, via the new mechanism, the proposed algorithm does not require any knowledge about the global and common sparsity structure of the involved channels, apart from the total number of non-zero AoDs. Also, it is noted that, to the best of the authors' knowledge, distributed sparse coding approaches (such as DiOMP) are for the first time employed for a channel estimation task. The proposed algorithm, called WDiOMP, consists of an initialization stage, an iterative stage, where the users collaborate to estimate the sparsity support sets, and, finally, a local channel estimation stage.
\subsection*{Algorithm WDiOMP} 
{\it First stage:
\begin{enumerate}
	\item For $T$ time slots, the BS transmits (broadcasts) a number of pilots, constituting matrix $\boldsymbol{X}$, to all users in the network.
	\item  Each user utilizes the collected measurements $\boldsymbol{y}_k$ and independently gets an initial estimation of its complete support set $\hat{\mathcal{S}}_k$, performing the standard OMP, initialized with an empty support set.
\end{enumerate}

Second stage (iterates over the following steps for $i=1$ to $L$, namely, up to the total sparsity level):
\begin{enumerate}
	\item User $k$, $\forall k$, transmits its $\hat{\mathcal{S}}_k$ to its neighboring users (using an appropriate D2D communication protocol \cite{jameel2018survey}), denoted by the set $\mathcal{L}_k^{out}$, and receives the estimates $\hat{\mathcal{S}}_l$'s from its neighbors, $\forall l \in \mathcal{L}_k^{in}$. 
	\item Each user utilizes the received $\hat{\mathcal{S}}_l$'s and applies a weighted majority voting mechanism to select the $i$ best indices.  
	\item The remaining non-zero indices, up to $L$, are acquired locally by each user by employing OMP.
\end{enumerate}

Third stage:\\
Each user utilizes the estimated set $\hat{\mathcal{S}}_k$ on \eqref{eq:ykn} by keeping only the relevant columns of $\boldsymbol{X\Psi}$ and finds the non-zero values by employing LS.}

The motivation for the weighted voting mechanism in the second step of the second stage of the Algorithm WDiOMP stems from the fact that a user may not contribute in the same manner to the support set recovery for several reasons like (a) different communication conditions (noise level), (b) different common support sets among groups and (c) lack of knowledge concerning the global and common parts of the support sets and the grouping of users. To cope with these cases, a weighted majority voting mechanism is employed which combines the votes of the cooperating users with a weight depicting the suitability of a user participating in a group. 
In more detail, at iteration $i$, user $k$ utilizes the vector $\boldsymbol{z}_{k,i}$ of size $\hat{L}$, initially, with zero elements.
For each received $\{\hat{\mathcal{S}}_{l,i}\}$, $l\in \mathcal{L}_k^{in}$, user $k$ adds the weight $a_{lk}$ to the positions of 
$\boldsymbol{z}_{k,i}$, indexed by $\hat{\mathcal{S}}_{l,i}$. User $k$, incorporating the information from all users $l\in\mathcal{L}_k^{in}$,
it subsequently selects the indexes of the $i$ largest elements of the final $\boldsymbol{z}_{k,i}$. The weight $a_{lk}$ is updated with a mechanism similar to the one in \cite{sayed2013diffusion} as follows. 
\begin{align}
b_{lk,i} & = (1-v)b_{lk,i-1} + v|\hat{\mathcal{S}}_{l,i} \setminus \hat{\mathcal{S}}_{k,i-1}|. \label{eq:b}\\
a_{lk,i} & = \frac{\frac{1}{b_{lk,i}}}{\sum_{m \in \mathcal{L}_k^{in}} \frac{1}{b_{mk,i}}}, l \in \mathcal{L}_k^{in}, k \in K \label{eq:a}.
\end{align}
In (\ref{eq:b}), the second term captures the number of different elements in the involved sets and $v \in (0,1)$ is a small positive factor. It is noted at this point that, for a sufficient number of measurements and high enough SNR values, when only global support set exists, it has been observed experimentally that the weights tend to $a_{lk} = \frac{1}{|\mathcal{L}_k^{in}|}$ for all users, while when both a global and common support sets exist, $a_{lk} = \frac{1}{|\mathcal{L}_k^{in}\cap\mathcal{C}_j|}$ for users in the same group and for users outside the group may tend to zero. 

Before concluding this section, the transmission efficiency in bits of the cooperative schemes, namely DiOMP and WDiOMP, is compared to the case when the measurements are gathered in a central point (for processing by a centralized algorithm like SOMP \cite{li2015decentralized}). Using $q$ bits for the real and imaginary parts of the measurements in $\boldsymbol{y}$ for feeding them back to the BS, $2\cdot K \cdot T \cdot q$ + $L \cdot \lceil{log_2(\hat{L})}\rceil$ bits are required, where the second term captures the transmission overhead of the recovered support sets indices back to the users. For the (W)DiOMP-based schemes, $K \cdot |\mathcal{L}_{k}^{out}| \cdot \lceil{log_2(\hat{L})}\rceil\cdot L^2$ bits are required during cooperation. For example, if $q=36$bits, $T=20$, $K=10$, $\hat{L}=200$, $L=5$, $|\mathcal{L}_k^{out}|=6$ $\forall k \in \mathcal{V}$, SOMP demands $14.4$ Kbits while (W)DiOMP-based schemes require $12$Kbits.

\section{Simulations}\label{sec:res}

In this section, the performance of WDiOMP will be presented under three experiments. $K = 10$ single antenna users are considered along with a BS with $N = 128$ antennas. An overcomplete, predefined dictionary $\boldsymbol{\Psi}$ (using DFT) is employed with $\hat{L}=200$ atoms, which define a corresponding grid for $[-\pi,\pi]$, while $\theta_l\in [-\pi,\pi]$,  and $T \in [10,40]$. The elements of the training matrix $\boldsymbol{X} $ are i.i.d. random variables that follow $\mathcal{CN}(0,\frac{1}{T})$ (as in \cite{zhang2018distributed}). The $L$ none-zero elements in $\boldsymbol{w}_k$ are also i.i.d. random variables following $\mathcal{N}(0,1)$ (as in \cite{wimalajeewa2014omp}). Each user has knowledge of only the total sparsity level $L$ (i.e., not the individual parameters $L_g$, $L_c$). Finally, it is assumed that all users are connected via a network and can share the positions of the non-zero elements in their $\boldsymbol{w}_k$'s, namely, their support sets. Thus, $\mathcal{L}_k^{in} = \mathcal{L}_k^{out} = \mathcal{V}$ holds.

The Average Support set Cardinality Error (ASCE) is considered for the evaluation. It takes values in the range $[0, 1]$ and is defined as follows
\begin{equation}
ASCE = 1- \frac{1}{K}\frac{1}{M}\sum_{m=1}^{M} \sum_{k=1}^{K}\bigg \{\frac{|\hat{\mathcal{S}}_k^{m} \cap \mathcal{S}_k|}{|\mathcal{S}_k|}\bigg\}.
\end{equation}
For the results, $M = 1000$ Monte Carlo simulations were performed. 

In the following, the performance of WDiOMP (using $v = 0.1$ for the calculation of the weights) will be compared with DiOMP (which employs a simple majority voting), the centralized algorithm SOMP \cite{li2015decentralized} and the per user (locally) employed OMP algorithm. Three scenarios will be presented and the $ASCE$ of the various approaches will be assessed versus the number of measurements $T$.

\begin{figure}
\centering
		\includegraphics[scale=0.5]{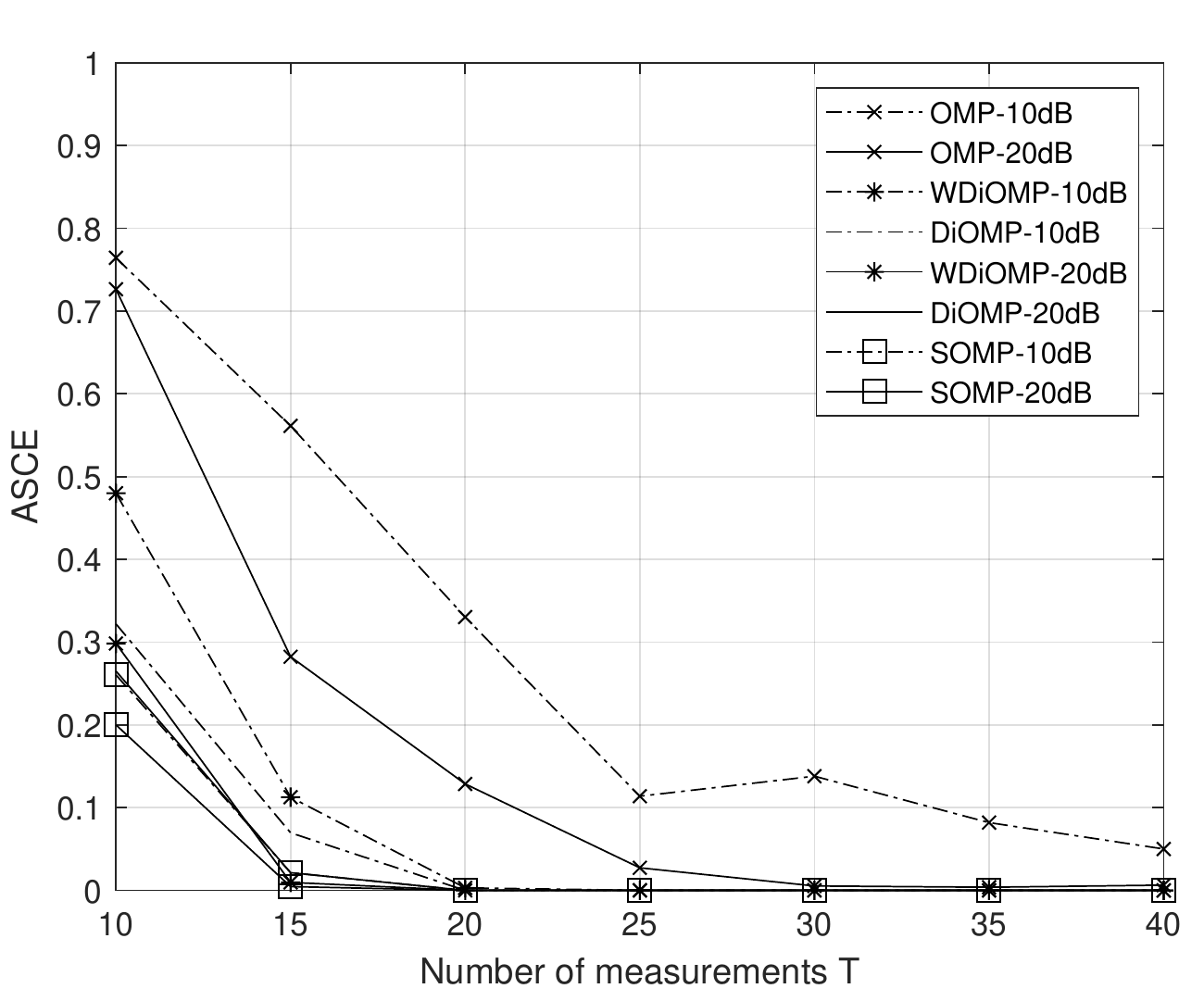}
		\caption{$L_g = 5$, $L_c=0$, $SNR = \{10, 20\}$dB}
		\label{fig:5a}
\end{figure}

In the first experiment, depicted in Fig.~\ref{fig:5a}, $ASCE$ is evaluated assuming that all users share a global support set and considering two $SNR$ values, namely, $10$ and $20$ dB. It is observed that both the DiOMP and WDiOMP achieve a performance close to the one achieved by SOMP, which,
being the point of reference for this experiment, was expected to demonstrate the best, though similar, performance. Additionally, the previous schemes perform
considerably better than the per-user (locally) employed OMP algorithm, resulting, for instance,  in about $40\%$ reduction in required training measurements for achieving near zero ASCE, for $SNR=20$ dB. Moreover, for $SNR=10$ dB, the reduction is even more pronounced as the OMP approach does not reach a near zero ASCE for the $T$ values considered. This experiment demonstrates the benefit of the users when they participate in a collaborative channel estimation procedure that exploits the adopted sparsity model, as opposed to the case in which each user operates in isolation. Finally, it should be mentioned that although DiOMP, WDiOMP and SOMP perform similarly, the first two do not utilize the BS during channel estimation. 

\begin{figure}
\centering
		\includegraphics[scale = 0.5]{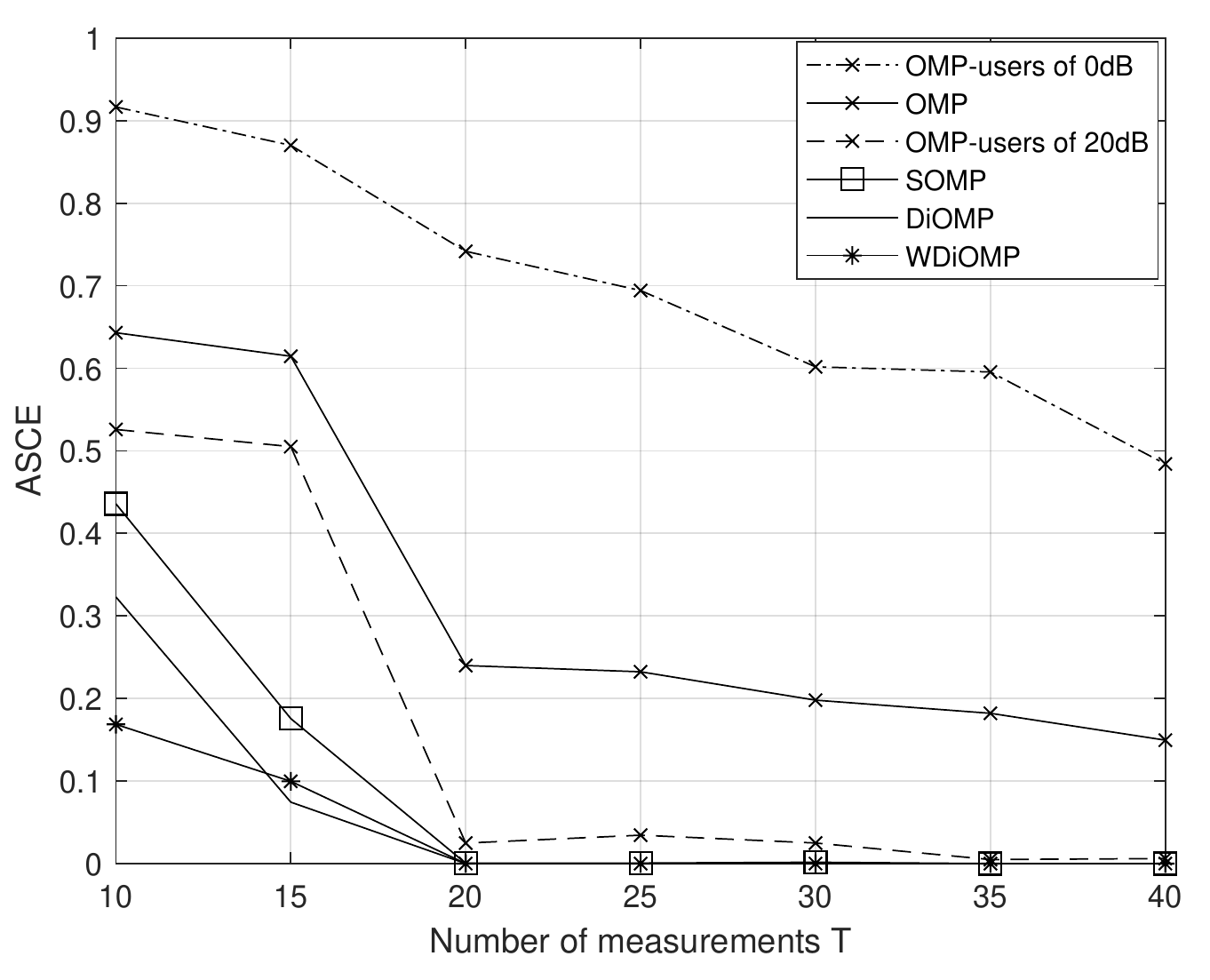}
		\caption{Different $SNR$ conditions per user. $L_g = 5$, $L_c=0$.}
		\label{fig:5b}
\end{figure}

In the second experiment, shown in Fig.~\ref{fig:5b}, an ``unbalanced'' scenario is examined in which, although the users share only a global support set, they operate in different SNR levels, namely, $SNR=20$ dB and $SNR=0$ dB for 7 and 3 users, respectively. 
For the OMP case, in which the users operate at isolation, apart from the overall mean performance (depicted by the ``OMP'' curve), the
performances when focusing on the groups that face $SNR=20$ dB and $SNR=0$ dB, are also presented. It is observed that the collaboration of the users (either centrally via SOMP or in a distributed fashion via DiOMP and WiOMP) improves
considerably the estimation performance over the case in which the users operate individually (via OMP) and, in particular, for users facing both the better and (especially) the worse conditions. Finally, it is observed that WDiOMP is able to start from a lower $ASCE$ which is attributed at its capability of weighting the contribution of each user, a behaviour that will be even more pronounced in the following experiment.  

\begin{figure}
\centering
		\includegraphics[scale=0.51]{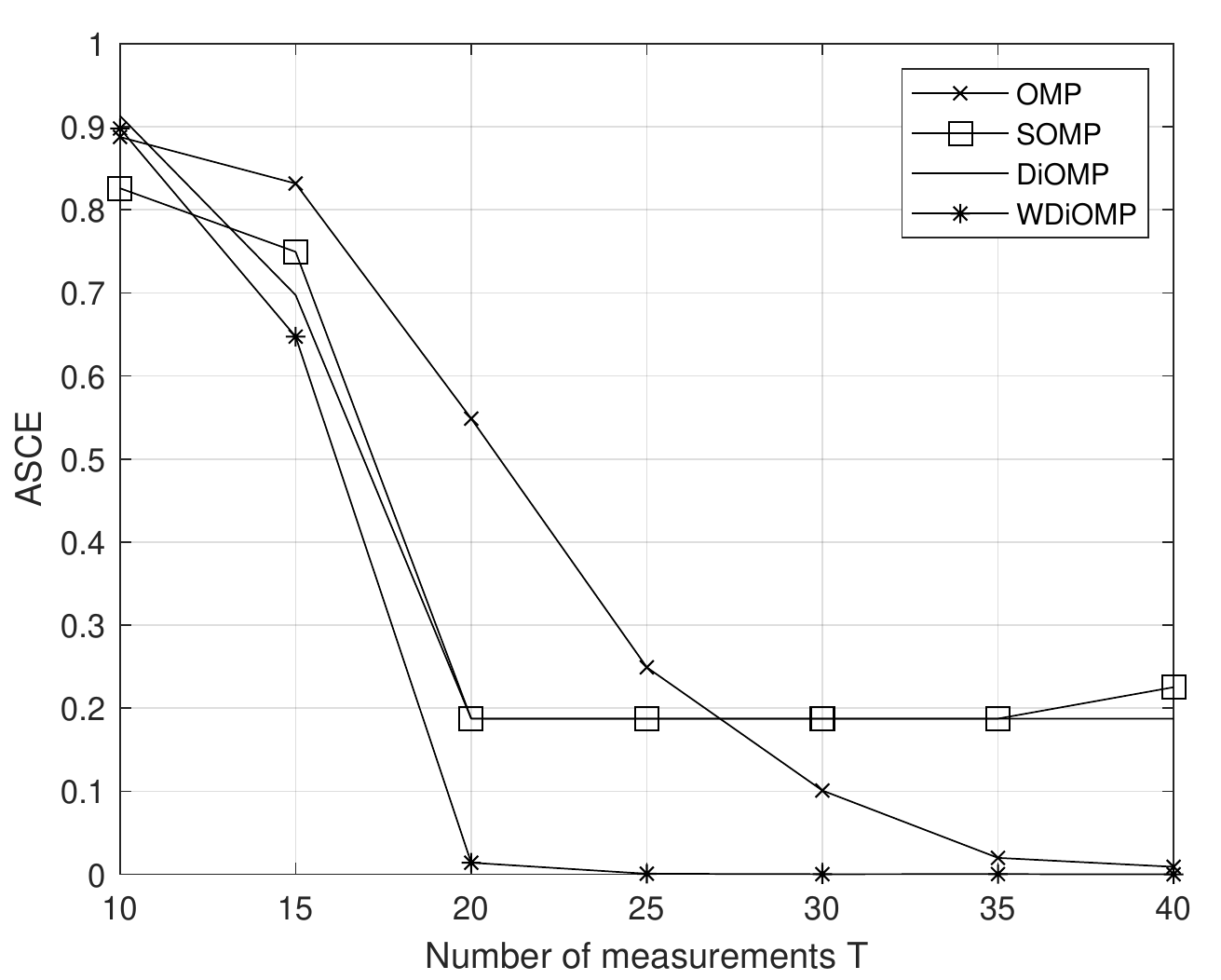}
		\caption{Two groups of users face global and common support sets. $L_g = 5$, $L_c=3$), $SNR = 20$dB.}
		\label{fig:6}
\end{figure}

In the third experiment (see Fig.~\ref{fig:6}), another ``unbalanced'' scenario is examined. In more detail, the users are divided into two groups that share a global support set (assuming $L_g=5$), while the users, of each group, share also a common support set of size $L_c=3$. Additionally, all users face $SNR=20$ dB. Let us recall that only the total sparsity level ($L=8$) is known (namely, $L_g$ and $L_c$ are considered unknown). 
It is observed that WDiOMP perfoms similarly to the previous experiments and considerably better than SOMP and DiOMP. For the latter, the floor in their performance is a result of the unknown $L_g$ and $L_c$. WiOMP does not have this issue because it is able to infer (in a wide sense) the sparsity structure via the employed weighting voting mechanism. Also, it is noted that SOMP, here, is shown in order to assess the performance of DiOMP and should not be considered as the centralized version of the examined scenario. Furthermore, WDiOMP is better than OMP and requires about $40\%$ less measurements for achieving a near-zero $ASCE$ (in the case of OMP no floor appears as only $L$ is relevant). Furthermore, it is mentioned that, although not shown here, the weights calculated by a particular user that employs WDiOMP, tend to similar (and larger) values for users in the same group while the weights are smaller for the remaining users, as hinted in Sec \ref{sec:CODIC}. As a final remark, it is noted that no $MSE$ curves for estimating $\boldsymbol{h}$ in (\ref{eq:channel}) are presented due to space limitations. However, the conclusions are similar to the ones for $ASCE$. From the previous results, WDiOMP demonstrates a robust performance for all considered experimental scenarios via the proposed weighting voting mechanism.
 
\section{Conclusion}
A fully distributed algorithm, WDiOMP, for the downlink channel estimation has been presented that exploits a general sparsity support model for the involved channels. The improved performance of WDiOMP has been assessed in various scenarios, even when the structure of the sparsity support is unknown. Our aim is to further investigate the entailed collaboration benefits in terms of communication and energy consumption.

\section{Acknowledgment}
This research is co-financed by Greece and the European Union 
via 'Strengthening Human Resources Research Potential via Doctorate Research' (MIS-5000432), implemented by the State Scholarships Foundation (IKY). This work is also supported by the ERDF and the Republic of Cyprus through the Project INFRASTRUCTURES/1216/0017
IRIDA.
\bibliographystyle{IEEEtran}
\bibliography{bibliography}
\end{document}